\documentclass[preprint,showpacs,aps,pre]{revtex4}
\usepackage{amsmath}
\usepackage{amssymb}
\usepackage{graphicx}
\usepackage{graphics}

\begin{document}

\title{Fluctuation effects in the theory of microphase separation of diblock
copolymers in the presence of an electric field}
\author{I. Gunkel, S. Stepanow, T. Thurn-Albrecht, S. Trimper}
\affiliation{Institut f\"{u}r Physik, Martin-Luther-Universit\"{a}t Halle-Wittenberg,
D-06099 Halle, Germany}
\date{\today}

\begin{abstract}
We generalize the Fredrickson-Helfand theory of the microphase separation in
symmetric diblock copolymer melts by taking into account the influence of a
time-independent homogeneous electric field on the composition fluctuations
within the self-consistent Hartree approximation. We predict that electric
fields suppress composition fluctuations, and consequently weaken the
first-order transition. In the presence of an electric field the critical
temperature of the order-disorder transition is shifted towards its
mean-field value. The collective structure factor in the disordered phase
becomes anisotropic in the presence of the electric field. Fluctuational
modulations of the order parameter along the field direction are strongest
suppressed. The latter is in accordance with the parallel orientation of the
lamellae in the ordered state.
\end{abstract}

\maketitle


\section{Introduction}

The effect of electric fields on the behavior of dielectric block copolymer
melts in bulk and in thin films has found increasing interest in recent
years \cite{wirtz-ma92}-\cite{matsen06} (and references therein) due to the
possibility to create uniform alignment in macroscopic microphase separated
samples. This is of special relevance for applications using self assembled
block copolymer structures for patterning and templating of nanostructures
\cite{park03}. The driving force for electric field induced alignment is the
orientation-dependent polarization in a material composed of domains with
anisotropic shape. The reason for the orientation in electric fields has a
very simple explanation in samples where the inhomogeneities appear only at
interfaces of cylinders or lamellae, as it is roughly the case in the strong
segregation limit. The polarization of the sample in this case induces
surface charges at the interfaces depending on the relative orientation of
the interfaces with respect to the field. The system is lowering its free
energy, if the interfaces orient parallel to the field. If the composition
of a block copolymer sample and consequently also the local dielectric
constant varies gradually, the polarization charges appear in the whole
system. However, interfaces parallel to the field possess the lowest
electric energy also in this case.

The effects of electric fields on the behavior of diblock copolymer melts
have been studied in \cite{AHQH+S94} by taking into account the electric
part to the free energy (quadratic in the strength of the electric field and
the order parameter) in addition to the thermodynamic potential including
composition fluctuations in the absence of the field. The influence of an
electric field on the composition fluctuations has not been considered yet.
However, the general relation between the derivatives of the thermodynamic
potential and the correlation function of the order parameter, given by Eq.(%
\ref{gf_phi}), requires the inclusion of the electric field into the
correlation function of the Brazovskii self-consistent Hartree approach,
too. The angular dependence of the structure factor without taking into
account the fluctuation effects was derived previously for polymer solutions
in \cite{wirtz-ma92}, and for copolymer melts in \cite{onuki-95}.
Intuitively it seems obvious that fluctuations become anisotropic in an
electric field, and moreover fluctuations of modes with wave vectors
parallel to the electric field are suppressed. The effects of an electric
field on composition fluctuations are directly accessible in scattering
experiments, and were studied for polymer solutions in \cite{wirtz-ma92} and
for asymmetric diblock copolymers in \cite{tta02}. In this paper we present
the results of the generalization of the Fredrickson-Helfand theory \cite%
{FrHe87} by taking into account the effects of the electric field on the
composition fluctuations in symmetric diblock copolymer melts.

The paper is organized as follows. Section \ref{sect-bulk} reviews the
collective description of the diblock copolymer melt. Section \ref%
{sect-efield} describes the coupling of the block copolymer melt to external
time-independent electric fields. Section \ref{sect-hartree} gives an
introduction to the Brazovskii-Fredrickson-Helfand treatment of composition
fluctuations in the presence of time-independent electric fields. Section %
\ref{sect-results} contains our results.

\section{Theory}

\subsection{A brief review of the collective description}

\label{sect-bulk}

\noindent In a diblock copolymer ($AB$), a chain of $N_{A}$ subunits of type
$A$ is at one end covalently bonded to a chain of $N_{B}$ subunits of type $%
B $. A net repulsive $A-B$ interaction energy $E\propto \varepsilon
_{AB}-(\varepsilon _{AA}+\varepsilon _{BB})/2$ between the monomers leads to
microphase separation. Thus at an order-disorder transition concentration
waves are formed spontaneously, having a wavelength of the same order as the
radius of gyration of the coils. The type of the long range order that forms
depends on the composition of the copolymers $f=N_{A}/N$ with $N=N_{A}+N_{B}$%
. Here we treat only the symmetric composition with $f=1/2$ for which the
lamellar mesophase is formed. As an order parameter we consider the
deviation of the density of $A$-polymers from its mean value, $\delta \Phi
_{A}(\mathbf{r)}=\rho _{A}(r)-f\rho _{m}$, where $\rho _{A}(r)$ is the
monomer density of the $A$ monomers and $\rho _{m}$ is the average monomer
density of the melt. Since the system is assumed to be incompressible, the
condition $\delta \Phi _{A}(\mathbf{r})+\delta \Phi _{B}(\mathbf{r})=0$
should be fulfilled. The expansion of the effective Landau Hamiltonian in
powers of the fluctuating order parameter $\delta \Phi (\mathbf{r})\equiv $ $%
\delta \Phi _{A}(\mathbf{r})$ was derived by Leibler \cite{leibler80} using
the random phase approximation (RPA). Following to Fredrickson and Helfand
\cite{FrHe87} let us introduce instead of $\delta \Phi (\mathbf{r})$ the
dimensionless order parameter $\psi (\mathbf{r)=}$ $\delta \Phi _{A}(\mathbf{%
r})/\rho _{m}$. The effective Hamiltonian in terms of $\psi (\mathbf{r)}$ is
given in units of $k_{B}T$ for symmetric composition by \cite{FrHe87}%
\begin{eqnarray}
H(\psi ) &=&\frac{1}{2}\int_{q}\psi (-\mathbf{q})\gamma _{2}(q)\psi (\mathbf{%
q})+\frac{1}{4!}\int_{q_{1}}\int_{q_{2}}\int_{q_{3}}\gamma _{4}(\mathbf{q}%
_{1},\mathbf{q}_{2},\mathbf{q}_{3},-\mathbf{q}_{1}-\mathbf{q}_{2}-\mathbf{q}%
_{3})  \notag \\
&\times &\psi (\mathbf{q}_{1})\psi (\mathbf{q}_{2})\psi (\mathbf{q}_{3})\psi
(-\mathbf{q}_{1}-\mathbf{q}_{2}-\mathbf{q}_{3})\,,  \label{1}
\end{eqnarray}%
where the wave vector $\mathbf{q}$ is the Fourier conjugate to $\mathbf{x}%
\equiv \rho _{m}^{1/3}\mathbf{r}$, and the shortcut $\int_{q}\equiv \int
\frac{d^{3}q}{(2\pi )^{3}}$ is introduced. Furthermore $\mathbf{q}_{r}$ is
the Fourier conjugate to $\mathbf{r}$ and is defined via $\mathbf{q}=$ $%
\mathbf{q}_{r}\rho _{m}^{-1/3}$. The quantity $\gamma _{2}(q)$ in Eq.~(\ref%
{1}) is defined as%
\begin{equation*}
\gamma _{2}(q)=\left( F(y)-2\chi N\right) /N
\end{equation*}%
with%
\begin{equation*}
F(y)=\frac{g_{D}(y,1)}{g_{D}(y,f)g_{D}(y,1-f)-\frac{1}{4}\left[
g_{D}(y,1)-g_{D}(y,f)-g_{D}(y,1-f)\right] ^{2}},
\end{equation*}%
and%
\begin{equation*}
g_{D}(y,f)=\frac{2}{y^{2}}\left( fy+e^{-fy}-1\right) \,.
\end{equation*}%
Here $g_{D}(y,f)$ is the Debye function and $y=q_{r}^{2}R_{g}^{2}=q^{2}\rho
_{m}^{2/3}R_{g}^{2}=q^{2}N/6$, where $R_{g}=$ $b\sqrt{N/6}$ is the
unperturbed radius of gyration of the copolymer chain. We assume that both
blocks have equal statistical segment lengths denoted by $b$ and furthermore
$b=\rho _{m}^{-1/3}$. The collective structure factor of the copolymer melt $%
S_{c}(q)=\left\langle \delta \Phi _{A}(\mathbf{q}_{r})\delta \Phi _{A}(-%
\mathbf{q}_{r})\right\rangle $ is related to $\gamma _{2}(q)$ by
\begin{equation*}
S_{c}(q)=\gamma _{2}^{-1}(q).
\end{equation*}%
The scattering function of a Gaussian polymer chain $S(q_{r})=(2/N)%
\sum_{i<j}\left\langle \exp (-i\mathbf{q_{r}r}_{ij}\right\rangle $ can be
expressed by the Debye function $g_{D}(x,1)$ via the relation $%
S(q)=Ng_{D}(x,1)$. The vertex function $\gamma _{4}(\mathbf{q}_{1},\mathbf{q}%
_{2},\mathbf{q}_{3},-\mathbf{q}_{1}-\mathbf{q}_{2}-\mathbf{q}_{3})$ is
expressed in the random phase approximation through the correlation
functions of one and two Gaussian copolymer chains \cite{leibler80}. The
vertex function as well as the quantity $F(y)$ in $\gamma _{2}(q)$ are
independent of temperature. The collective structure factor $S_{c}(q)$ has a
pronounced peak at the wave vector $q^{\ast }$, obeying the condition $%
y^{\ast }=\left( q_{r}^{\ast }R_{g}\right) ^{2}=3.7852$ for $f=1/2$ , i.e. $%
y^{\ast }$ is independent of both temperature and molecular weight. As it is
well-known \cite{FrHe87}, the composition fluctuations, which are described
according to Brazovskii \cite{brazovskii75} within the self-consistent
Hartree approach, change the type of the phase transition to the ordered
state from second order to fluctuation-induced weak first-order. In this
theory \cite{brazovskii75}, \cite{FrHe87} the inverse of the collective
structure factor is approximated near to the peak by%
\begin{eqnarray}
N\gamma _{2}(q) &=&F(y^{\ast })-2\chi N+\frac{1}{2}\frac{\partial ^{2}F}{%
\partial y^{\ast 2}}\left( y-y^{\ast }\right) ^{2}+...  \notag \\
&=&2\left( \chi N\right) _{s}-2\chi N+\frac{1}{2}\frac{\partial ^{2}F}{%
\partial y^{\ast 2}}4y^{\ast }\frac{N}{6}\left( q-q^{\ast }\right) ^{2}+...
\notag \\
&\simeq &Nc^{2}\left( \tilde{\tau}+\left( q-q^{\ast }\right) ^{2}\right) .
\label{S_c1}
\end{eqnarray}%
According to \cite{FrHe87} the notations%
\begin{eqnarray}
\left( 2\chi N\right) _{s} &=&F(y^{\ast })=20.990,  \notag \\
c &=&\sqrt{y^{\ast }\partial ^{2}F/\partial y^{\ast 2}/3}=1.1019,  \notag \\
\tilde{\tau} &=&\frac{2\left( \chi N\right) _{s}-2\chi N}{Nc^{2}}
\label{tildtau}
\end{eqnarray}%
are introduced. Redefining the order parameter by $\psi =c^{-1}\phi $ and
inserting Eq.~(\ref{S_c1}) into Eq.~(\ref{1}), the effective Hamiltonian can
be written as%
\begin{eqnarray}
H(\phi ) &=&\frac{1}{2}\int_{q}\phi (-\mathbf{q})\left( \tilde{\tau}+\left(
q-q^{\ast }\right) ^{2}\right) \phi (\mathbf{q})  \notag \\
&+&\frac{\tilde{\lambda}}{4!}\int_{q_{1}}\int_{q_{2}}\int_{q_{3}}\phi (%
\mathbf{q}_{1})\phi (\mathbf{q}_{2})\phi (\mathbf{q}_{3})\phi (-\mathbf{q}%
_{1}-\mathbf{q}_{2}-\mathbf{q}_{3}),  \label{H_Phi}
\end{eqnarray}%
where $\tilde{\lambda}=\gamma _{4}(\mathbf{q}^{\ast },\mathbf{q}^{\ast },%
\mathbf{q}^{\ast },-\mathbf{q}^{\ast }-\mathbf{q}^{\ast }-\mathbf{q}^{\ast
})/c^{4}$. Following \cite{FrHe87} the vertex $\gamma _{4}$\ is approximated
by its value at $\mathbf{q}^{\ast }$. The quantity $\tilde{\tau}$ plays the
role of the reduced temperature in the Landau theory of phase transitions.
Note that the scattering function is obtained from Eq.~(\ref{S_c1}) as%
\begin{equation}
S_{c}^{-1}(q)=\tilde{\tau}+\left( q-q^{\ast }\right) ^{2}.  \label{scat-fl}
\end{equation}

In order to study the composition fluctuations in symmetric diblock
copolymer melts based on the effective Hamiltonian (\ref{H_Phi}), it is
convenient to introduce an auxiliary field $h(\mathbf{x})$, which is coupled
linearly to the order parameter. Thus, the term $\int d^{3}rh(\mathbf{x}%
)\phi (\mathbf{x})$ should be added to Eq.~(\ref{H_Phi}). Consequently the
average value of the order parameter can be written as $\bar{\phi}(\mathbf{x}%
)=-\delta \mathcal{F}(h)/\delta h(\mathbf{x})|_{h=0}$, where $\mathcal{F}(h)$
is the free energy related to the partition function $Z(h)$ by $\mathcal{F}%
(h)=-k_{B}T\ln Z(h)$. The Legendre transformation, $\delta \left( \mathcal{F}%
+\int d^{3}rh(\mathbf{x})\bar{\phi}(\mathbf{x})\right) \equiv \delta \Gamma (%
\bar{\phi})=\int d^{3}rh(\mathbf{x})\delta \bar{\phi}(\mathbf{x})$
introduces the thermodynamic potential $\Gamma (\bar{\phi})$, which is a
functional of the average value of the order parameter $\bar{\phi}(\mathbf{x}%
)$. In terms of the Gibbs potential $\Gamma (\bar{\phi})$ the spontaneous
value of the order parameter is determined by the equation $0=\delta \Gamma (%
\bar{\phi})/\delta \bar{\phi}(\mathbf{x})$. The potential $\Gamma (\bar{\phi}%
)$ is the generating functional of the one-particle irreducible Greens's
function, and can be represented as a series by using Feynman diagrams \cite%
{zinn-justin}. The second derivative of the Gibbs potential with respect to
the order parameter yields the inverse correlation function%
\begin{equation}
\frac{\delta ^{2}\Gamma (\bar{\phi})}{\delta \bar{\phi}(\mathbf{x}%
_{1})\delta \bar{\phi}(\mathbf{x}_{2})}=S^{-1}(\mathbf{x}_{1}\mathbf{,x}_{2},%
\bar{\phi}).  \label{gf_phi}
\end{equation}%
The correlation functions of composition fluctuations in the disordered and
ordered phase are defined by
\begin{equation*}
S(\mathbf{x}_{1}\mathbf{,x}_{2})=\left\langle \phi (\mathbf{x}_{1})\phi (%
\mathbf{x}_{2})\right\rangle ,\ \ S(\mathbf{x}_{1}\mathbf{,x}_{2},\bar{\phi}%
)=\left\langle \Delta \phi (\mathbf{x}_{1})\Delta \phi (\mathbf{x}%
_{2})\right\rangle ,
\end{equation*}%
respectively, where the abbreviation $\Delta \phi (\mathbf{x})=\phi (\mathbf{%
x})-\bar{\phi}(\mathbf{x})$ is introduced. Eq.~(\ref{gf_phi}) is the pendant
of the well-known relation
\begin{equation}
\frac{\delta ^{2}\mathcal{F}(h)}{\delta h(\mathbf{x}_{1})\delta h(\mathbf{x}%
_{2})}|_{\,h=0}=S(\mathbf{x}_{1}\mathbf{,x}_{2}),  \label{F-diff2}
\end{equation}%
which represents the relation between the thermodynamic quantities and the
correlation function of the composition $\phi (\mathbf{x})$. For $\bar{\phi}%
=0$ Eq.~(\ref{gf_phi}) is obtained from Eq.~(\ref{F-diff2}) using the
Legendre transform. In case of a constant auxiliary field the variational
derivatives on the left-hand side of Eq.~(\ref{F-diff2}) should be replaced
by the partial ones, while the integration over $\mathbf{x}_{1}$\textbf{\ }%
and\textbf{\ }$\mathbf{x}_{2}$ is carried out on the right-hand side. Then
the derivative $-\partial \mathcal{F}/\partial h$ is the mean value of the
order parameter $\bar{\phi}(\mathbf{x})$ multiplied by the volume of the
system. The 2nd derivative of the free energy, $\partial ^{2}\mathcal{F}%
/\partial h^{2}$, is related to the susceptibility, which in accordance to
Eq.~(\ref{F-diff2}) is equal to the integral of the correlation function.

The order parameter for a symmetric diblock copolymer melt can be
approximated in the vicinity of the critical temperature of the microphase
separation by
\begin{equation}
\bar{\phi}(\mathbf{x})=2A\cos \left( q^{\ast }\mathbf{nx}\right) ,
\label{OP}
\end{equation}%
where $\mathbf{n}$ is an unit vector in the direction of the wave vector
perpendicular to the lamellae and $A$ is an amplitude. The Brazovskii
self-consistent Hartree approach, which takes into account the fluctuation
effects on the microphase separation, is based on the following expression
of the derivative of the Gibbs potential with respect to the amplitude $A$
of the order parameter%
\begin{equation}
\frac{1}{A}\frac{\partial \Gamma (A)/V}{\partial A}=2\left( \tilde{\tau}+%
\frac{\tilde{\lambda}}{2}\int_{q}S_{0}(\mathbf{q,}A)+\frac{\tilde{\lambda}}{2%
}A^{2}\right) .  \label{braz1}
\end{equation}%
The second term in Eq.~(\ref{braz1}) includes the propagator
\begin{equation}
S_{0}(\mathbf{q,}A)=1/\left( \tilde{\tau}+\left( q-q^{\ast }\right) ^{2}+%
\tilde{\lambda}A^{2}\right)  \label{2}
\end{equation}%
and represents the first-order correction to the thermodynamic potential
owing to the self-energy. The first two terms in the brackets on the
right-hand side of Eq.~(\ref{braz1}) are summarized to an effective reduced
temperature denoted by $\tilde{\tau}_{r}$. The equation for $\tilde{\tau}%
_{r} $ becomes self-consistent by replacing $\tilde{\tau}$ in Eq.~(\ref{2})
for $S_{0}(\mathbf{q,}A)$ by $\tilde{\tau}_{r}$. Then we get
\begin{equation}
\tilde{\tau}_{r}=\tilde{\tau}+\frac{\tilde{\lambda}}{2}\int_{q}S(\mathbf{q,}%
A),  \label{braz2}
\end{equation}%
where $S^{-1}(\mathbf{q,}A)=\tilde{\tau}_{r}+\left( q-q^{\ast }\right) ^{2}+%
\tilde{\lambda}A^{2}$ is the inverse of the effective propagator.

\subsection{Contribution of the electric field to the effective Hamiltonian}

\label{sect-efield}

\noindent In this subsection we discuss the coupling of the diblock
copolymer melt to an external time-independent electric field. \ The system
we consider is a linear dielectric and is free of external charges. The
field satisfies the Maxwell equation%
\begin{equation*}
\mathrm{div}\left( \varepsilon (\mathbf{r})\mathbf{E}(\mathbf{r})\right) =0,
\end{equation*}%
where the inhomogeneities of the dielectric constant $\varepsilon (\mathbf{r}%
)$ are caused by the inhomogeneities of the order parameter. According to
\cite{AHQS93} we adopt the expansion of the dielectric constant in powers of
the order parameter up to quadratic terms%
\begin{equation}
\varepsilon (\mathbf{r})=\varepsilon _{D}(\mathbf{r})+\beta \bar{\phi}(%
\mathbf{r})+\frac{1}{2}\frac{\partial ^{2}\varepsilon }{\partial \bar{\phi}%
^{2}}\bar{\phi}(\mathbf{r})^{2}.  \label{eps-r}
\end{equation}%
In case of zero order parameter $\bar{\phi}(\mathbf{r})=0$ the dielectric
constant is assumed to be homogeneous, i.e. $\varepsilon _{D}(\mathbf{r}%
)=\varepsilon _{D}$. The above Maxwell equation can be rewritten as integral
equation as follows%
\begin{equation}
\mathbf{E}(\mathbf{r})=\mathbf{E}_{0}+\frac{1}{4\pi }\boldsymbol{\nabla }%
\int d^{3}r_{1}G_{0}(\mathbf{r}-\mathbf{r}_{1})\left( \mathbf{E}(\mathbf{r}%
_{1})\boldsymbol{\nabla }\right) \ln \varepsilon (\mathbf{r}_{1}),
\label{E_r}
\end{equation}%
where $G_{0}(\mathbf{r})=1/r$ is the Green's function of the Poisson
equation. The integral equation (\ref{E_r}) is convenient to derive
iterative solutions for the electric field, and to take into account the
dependencies of $\varepsilon (r)$\ on the order parameter.\ The 2nd term in
Eq.~(\ref{E_r}) takes into account the polarization due to the
inhomogeneities of the order parameter. The substitution $\mathbf{E}(\mathbf{%
r}_{1})=\mathbf{E}_{0}$ on the right-hand side of Eq.~(\ref{E_r}) gives the
first-order correction to the external electric field as%
\begin{eqnarray}
\mathbf{E}(\mathbf{r}) &=&\mathbf{E}_{0}+\mathbf{E}_{1}(\mathbf{r})+\dots
\notag \\
&=&\mathbf{E}_{0}+\frac{1}{4\pi }\frac{\beta }{\varepsilon _{D}}%
\boldsymbol{\nabla }\int d^{3}r_{1}G_{0}(\mathbf{r}-\mathbf{r}_{1})\left(
\mathbf{E}_{0}\boldsymbol{\nabla }\right) \bar{\phi}(\mathbf{r}_{1})+\dots
\label{E-r1}
\end{eqnarray}%
The higher-order terms $\mathbf{E}_{i}(\mathbf{r})$ ($i=2$, $3$,$...$) in
the last equation are linear in the external field, too.

In taking into account the electric energy in thermodynamic potentials one
should distinguish between the thermodynamic potentials with respect to the
charges or the potential \cite{landau-lifshitz8}. These thermodynamic
potentials are connected with each other by a Legendre transformation. Here,
in calculating the effects of fluctuations we interpret in fact the Landau
free energy as an Hamiltonian, which weights the fluctuations by the
Boltzmann factor $\exp (-H)$. Therefore, the contribution of the electric
field to the effective Hamiltonian corresponds to the energy of the electric
field, and is given in Gaussian units by%
\begin{equation*}
k_{B}T\Gamma _{el}=\frac{1}{8\pi }\int d^{3}r\varepsilon (\mathbf{r})\mathbf{%
E}^{2}(\mathbf{r}),
\end{equation*}%
where $\varepsilon (\mathbf{r})$ and $\mathbf{E}(\mathbf{r})$ are given by
Eqs.~(\ref{eps-r},\ref{E-r1}).

In the following we consider only the polarization part of $\Gamma _{el}$.
The quadratic part of the latter in powers of the order parameter is given
by
\begin{eqnarray}
k_{B}T\Gamma _{el} &=&\frac{1}{8\pi }\int d^{3}r\frac{1}{2}\frac{\partial
^{2}\varepsilon }{\partial \bar{\phi}^{2}}\bar{\phi}(\mathbf{r})^{2}\mathbf{E%
}_{0}^{2}  \notag \\
&&+\frac{1}{8\pi }\int d^{3}r\varepsilon _{D}\frac{1}{4\pi }\nabla ^{m}\int
d^{3}r_{1}G_{0}(\mathbf{r}-\mathbf{r}_{1})E_{0}^{n}\frac{\beta }{\varepsilon
_{D}}\nabla ^{n}\bar{\phi}(\mathbf{r}_{1})  \notag \\
&&\times \frac{1}{4\pi }\nabla ^{m}\int d^{3}r_{2}G_{0}(\mathbf{r}-\mathbf{r}%
_{2})E_{0}^{k}\frac{\beta }{\varepsilon _{D}}\nabla ^{k}\bar{\phi}(\mathbf{r}%
_{2})  \notag \\
&=&\frac{1}{2}\int d^{3}r_{1}\int d^{3}r_{2}\bar{\phi}(\mathbf{r}_{1})\tilde{%
\gamma}_{2}^{el}(\mathbf{r}_{1}\mathbf{,r}_{2})\bar{\phi}(\mathbf{r}_{2}),
\label{G_el}
\end{eqnarray}%
with%
\begin{equation*}
\tilde{\gamma}_{2}^{el}(\mathbf{r}_{1}\mathbf{,r}_{2})=\frac{1}{8\pi }\frac{%
\partial ^{2}\varepsilon }{\partial \bar{\phi}^{2}}\mathbf{E}_{0}^{2}\delta (%
\mathbf{r}_{1}\mathbf{-r}_{2})+\frac{\beta ^{2}}{\left( 4\pi \right)
^{3}\varepsilon _{D}}\int d^{3}r\nabla _{\mathbf{r}}^{m}\nabla _{\mathbf{r}%
_{1}}^{n}G_{0}(\mathbf{r}-\mathbf{r}_{1})\nabla _{\mathbf{r}}^{m}\nabla _{%
\mathbf{r}_{2}}^{k}G_{0}(\mathbf{r}-\mathbf{r}_{2})E_{0}^{n}E_{0}^{k}.
\end{equation*}%
Notice the sum convention over the indices $m$, $n$, $k$\ (= x, y, z) in the
above two equations and in Eq.~(\ref{CF-el}). Expressing $\Gamma _{el}$ by
the Fourier components of the order parameter yields
\begin{equation}
\Gamma _{el}=\frac{1}{2}\int_{q}\bar{\phi}(-\mathbf{q})\tilde{\gamma}%
_{2}^{el}(\mathbf{q})\bar{\phi}(\mathbf{q}),  \label{G_el-FT}
\end{equation}%
whereas $\tilde{\gamma}_{2}^{el}(\mathbf{q})$ is given by
\begin{equation}
\tilde{\gamma}_{2}^{el}(\mathbf{q})=\frac{1}{4\pi \rho _{m}k_{B}T}\left(
\frac{1}{2}\frac{\partial ^{2}\varepsilon }{\partial \bar{\phi}^{2}}\mathbf{E%
}_{0}^{2}+\frac{\beta ^{2}}{\varepsilon _{D}}\frac{q^{n}q^{k}}{\mathbf{q}^{2}%
}E_{0}^{n}E_{0}^{k}\right) .  \label{CF-el}
\end{equation}%
The electric contribution to the correlation function can be obtained using
Eq.~(\ref{gf_phi}). Note that the factor $\rho _{m}^{-1}$ in Eq.~(\ref{CF-el}%
) is due to the length redefinition $\mathbf{r}=\rho _{m}^{-1/3}\mathbf{x}$.
Eqs.~(\ref{G_el})-(\ref{CF-el}) are used in the subsequent section to
analyze the influence of the electric field on the composition fluctuations
of the order parameter. Assuming that the electric field is directed along
the $z$-axes, and denoting the angle between the field and the wave vector $%
\mathbf{q}$ by $\theta $, we obtain the quantity $\tilde{\gamma}_{2}^{el}(%
\mathbf{q})$ in Eq.~(\ref{CF-el}) as
\begin{equation}
\tilde{\gamma}_{2}^{el}(\mathbf{q})=\tilde{\alpha}\cos ^{2}\theta +\tilde{%
\alpha}_{2},  \label{gamma-el}
\end{equation}%
where the notations
\begin{equation}
\tilde{\alpha}=\frac{1}{4\pi \rho _{m}k_{B}T}\frac{\beta ^{2}}{\varepsilon
_{D}}\mathbf{E}_{0}^{2},\ \ \tilde{\alpha}_{2}=\frac{1}{4\pi \rho _{m}k_{B}T}%
\frac{1}{2}\frac{\partial ^{2}\varepsilon }{\partial \bar{\phi}^{2}}\mathbf{E%
}_{0}^{2}  \label{alpha}
\end{equation}%
are used.

The thermodynamic potential given by Eq.~(\ref{G_el}) is quadratic in both
the external electric field $E_{0}$\ and the order parameter. The terms $%
E_{i}(r)$ with $i>1$ in Eq.~(\ref{E-r1}) are still linear in $E_{0}$, but
contain higher powers of the order parameter and its derivatives. In the
vicinity of the order-disorder transition, where the order parameter is
small and its inhomogeneities are weak, the higher-order contributions to $%
\Gamma _{el}$\ can be neglected.

\subsection{Hartree treatment of fluctuations in the presence of the
electric field}

\label{sect-hartree}

The Brazovskii-Hartree approach in absence of the electric field is
summarized in Eqs.~(\ref{braz1}, \ref{braz2}). According to Eq.~(\ref{gf_phi}%
) the contribution of the electric field should be taken into account both
to the thermodynamic potential and the propagator. Therefore, instead of
Eq.~(\ref{braz1}) we obtain

\begin{eqnarray}
&&\frac{1}{A}\frac{\partial }{\partial A}\left( g-\tilde{\gamma}_{2}^{el}(%
\mathbf{q}^{\ast })A^{2}\right) =\tilde{\tau}+  \notag \\
&&2\tilde{\lambda}\int_{q}\frac{1}{\tilde{\tau}+\tilde{\gamma}_{2}^{el}(%
\mathbf{q})+(q-q^{\ast })^{2}+\tilde{\lambda}\,A^{2}}+\tilde{\lambda}\,A^{2},
\label{GAns}
\end{eqnarray}%
where $g=\Gamma (A)/V$ is the thermodynamic potential per volume. The term $%
\tilde{\gamma}_{2}^{el}(q^{\ast })$\ in (\ref{GAns}) is the contribution to $%
g$ associated with the lamellae in the electric field below the transition
with the orientation defined by the angle between the wave vector $\mathbf{q}%
^{\ast }$ and the field strength $\mathbf{E}_{0}$, $\mathbf{q}^{\ast }%
\mathbf{E}_{0}=q^{\ast }E_{0}\cos \theta ^{\ast }$. The equilibrium
orientation of the lamellae is derived by minimization of the thermodynamic
potential with respect to the angle $\theta ^{\ast }$, and yields $\theta
^{\ast }=\pi /2$. As a consequence, the modulations of the order parameter
perpendicular to the electric field possess the lowest electric energy. The
term $\tilde{\gamma}_{2}^{el}(\mathbf{q}^{\ast })A^{2}$ in (\ref{GAns})
disappears in equilibrium state and will not be considered below. The
isotropic part of $\tilde{\gamma}_{2}^{el}(\mathbf{q})$ in Eq.~(\ref{alpha}%
), which is associated with $\tilde{\alpha}_{2}$, shifts $\tilde{\tau}$ for\
positive $\partial ^{2}\varepsilon /\partial \bar{\phi}^{2}$ to higher
values i.e. shifts the critical temperature to lower values and favors
mixing. The sign of this term agrees with that in \cite{wirtz-ma92}, where
the effect of this term was observed and studied for polymer solutions near
the critical point. Since this term is isotropic and small, it will not
contribute directly to alignment and will be not considered further. The
demixing of the low molecular mixtures studied recently in \cite%
{tsori-nature04}\ is due to field gradients. The fluctuations in the
presence of the electric field become anisotropic due to $\tilde{\gamma}%
_{2}^{el}(\mathbf{q})$ in Eq.~(\ref{GAns}). The first two terms on the
right-hand side of Eq.~(\ref{GAns}) define as before an effective $\tilde{%
\tau}$, which is denoted by $\tilde{\tau}_{r}$. Replacing $\tilde{\tau}$
under the integral in Eq.~(\ref{GAns}) by $\tilde{\tau}_{r}$ we obtain a
self-consistent equation for $\tilde{\tau}_{r}$
\begin{equation}
\tilde{\tau}_{r}=\tilde{\tau}+\frac{\tilde{\lambda}}{2}\int_{q}\frac{1}{%
\tilde{\tau}_{r}+\tilde{\alpha}\cos ^{2}\theta +(q-q^{\ast })^{2}+\tilde{%
\lambda}\,A^{2}}\,,  \label{m}
\end{equation}%
which generalizes Eq.~(\ref{braz2}) for $\mathbf{E}\neq 0$. Carrying out the
integration over $\mathbf{q}$ we obtain
\begin{eqnarray}
\frac{\tilde{\lambda}}{2}\int_{q}\frac{1}{\tilde{\tau}_{r}+\tilde{\lambda}%
\,A^{2}+\tilde{\alpha}\cos ^{2}\theta +(q-q^{\ast })^{2}} &=&\frac{\tilde{%
\lambda}q_{\ast }^{2}}{4\pi \sqrt{\tilde{\alpha}}}\mathrm{arcsinh}\sqrt{%
\frac{\tilde{\alpha}}{\tilde{\tau}_{r}+\tilde{\lambda}\,A^{2}}}  \notag \\
&=&\frac{d\tilde{\lambda}}{N\sqrt{\tilde{\alpha}}}\mathrm{arcsinh}\sqrt{%
\frac{\tilde{\alpha}}{\tilde{\tau}_{r}+\tilde{\lambda}\,A^{2}}},
\label{HarApp}
\end{eqnarray}%
where the notation $d=3y^{\ast }/2\pi $ with $y^{\ast }=q^{\ast
2}\,N/6=3.7852$ is used. Eq.~(\ref{m}) shows that the fluctuations in the
presence of the electric field are suppressed due to the angular dependence
of the integrand. Consequently an electric field weakens the first-order
phase transition. In computing the integral in Eq.~(\ref{HarApp}) we realize
that the leading contribution comes from the peak of the structure factor at
$q^{\ast }$. The contributions of large wave vectors, which become finite
after an introduction of an appropriate cutoff at large $q$, are expected to
renormalize the local parameters such as the $\chi $ parameter etc. Very
recently these fluctuations have been considered in \cite{kudlay03}. The
self-consistent equation for $\tilde{\tau}_{r}$ in the presence of the
electric field reads
\begin{equation}
\tilde{\tau}_{r}=\tilde{\tau}+\frac{\,\tilde{\lambda}d}{N\sqrt{\tilde{\alpha}%
}}\mathrm{arcsinh}\sqrt{\frac{\tilde{\alpha}}{\tilde{\tau}_{r}+\tilde{\lambda%
}\,A^{2}}}\,.  \label{tau-E}
\end{equation}%
The equation for $\tilde{\tau}_{r}$ in the Brazovskii-Fredrickson-Helfand
theory is obtained from Eq.~(\ref{tau-E}) in the limit $\alpha \rightarrow 0$
as
\begin{equation*}
\tilde{\tau}_{r}=\tilde{\tau}+\frac{\,\tilde{\lambda}d}{N}\left( \tilde{\tau}%
_{r}+\tilde{\lambda}A^{2}\right) ^{-1/2}\,.
\end{equation*}%
Because the quantities $\tilde{\tau}$ and $\tilde{\lambda}$ are of the order
$O(N^{-1})$, it is convenient to replace them in favor of
\begin{equation}
\tau =\tilde{\tau}N\ ,\ \lambda =\tilde{\lambda}N.  \label{tau}
\end{equation}%
Note that the transformation from $\tilde{\tau}$ to $\tau $ implies the
redefinition of the thermodynamic potential $g\rightarrow gN$. Instead of
Eq.~(\ref{tau-E}) we then obtain%
\begin{equation}
\tau _{r}=\tau +\frac{\,\lambda d}{\sqrt{N\alpha }}\mathrm{arcsinh}\sqrt{%
\frac{\alpha }{\tau _{r}+\lambda \,A^{2}}}\,,  \label{tau-E1}
\end{equation}%
where the quantity $\alpha $ is defined by $\alpha =\tilde{\alpha}N=(\beta
^{2}N/4\pi \rho _{m}k_{B}T\varepsilon _{D})\mathbf{E}_{0}^{2}$. For
symmetric composition $\lambda $ was computed in \cite{leibler80} as $%
\lambda =106.18$. Using the substitution $t=\tau _{r}+\lambda \,A^{2}$ we
obtain
\begin{equation}
t=\tau +\frac{\,\lambda d}{\sqrt{N\alpha }}\mathrm{arcsinh}\sqrt{\frac{%
\alpha }{t}}+\lambda \,A^{2}.  \label{t}
\end{equation}%
The derivative of the potential $g$ with respect to the amplitude of the
order parameter $A$ is obtained from Eq.~(\ref{GAns}) as
\begin{equation*}
\frac{\partial g}{\partial A}=2\tau _{r}(A)\,A+\lambda
\,A^{3}=2t(A)\,A-\lambda \,A^{3}.
\end{equation*}%
The integration gives the thermodynamic potential as
\begin{equation}
g=\int\limits_{0}^{A}2t(A)\,A\,dA-\frac{\lambda }{4}A^{4}=\frac{1}{2\lambda }%
\left( t{}^{2}-t_{0}^{2}\right) +\frac{d}{\sqrt{N}}\left( \sqrt{t+\alpha }-%
\sqrt{t_{0}+\alpha }\right) -\frac{\lambda }{4}A^{4}\,,  \label{g}
\end{equation}%
whereas the inverse susceptibility of the disordered phase $t_{0}\equiv
t(A=0)$ satisfies the equation%
\begin{equation}
t_{0}=\tau +\frac{d\,\lambda }{\sqrt{N\alpha }}\mathrm{arcsinh}\sqrt{\frac{%
\alpha }{t_{0}}}.  \label{t0}
\end{equation}%
Eqs.~(\ref{t}, \ref{g}, \ref{t0}) generalize the Fredrickson-Helfand
treatment of the composition fluctuations in symmetric diblock copolymer
melts in the presence of an external time-independent electric field.

\section{Results}

\label{sect-results}

\noindent The position of the phase transition is determined by the
conditions%
\begin{equation}
g=0,\ \ \frac{\partial g}{\partial A}=0,  \label{T_c}
\end{equation}%
and result in the following equation
\begin{equation}
\frac{1}{2}\left( t{}^{2}+t_{0}^{2}\right) =\frac{d\lambda }{\sqrt{N}}\left(
\sqrt{t+\alpha }-\sqrt{t_{0}+\alpha }\right) .  \label{gTr}
\end{equation}%
The perturbational solution of Eqs.~(\ref{t}), (\ref{t0}) and (\ref{gTr}) to
the first order in powers of $\alpha $ yields for the transition temperature
\begin{equation}
\tau _{t}=-2.0308\left( d\,\lambda \right) ^{2/3}N^{-1/3}+0.48930\alpha .
\label{tau_t}
\end{equation}%
By making use of Eqs.~(\ref{tildtau}) and (\ref{tau}) we find
\begin{equation}
\left( \chi N\right) _{t}=\left( \chi N\right) _{s}+1.0154\,c^{2}\left(
d\,\lambda \right) ^{2/3}N^{-1/3}-0.48930\alpha \frac{c^{2}}{2}.
\label{chi_t}
\end{equation}%
The solution of Eqs.~(\ref{t}) and (\ref{gTr}) results in the following
expression for the amplitude of the order parameter at the transition
\begin{equation}
A_{t}=1.4554\,\left( d^{2}/\lambda \right) ^{1/6}\,N^{-1/6}-0.42701\,\left(
d^{2}\lambda ^{5}\right) ^{-1/6}\,N^{1/6}\,\alpha \,,  \label{A}
\end{equation}%
where $\alpha $\ is defined by $\alpha =N\beta ^{2}/(4\pi \rho
_{m}k_{B}T\varepsilon _{D})\mathbf{E}_{0}^{2}$. The corrections associated
with the electric field in Eqs.~(\ref{tau_t}-\ref{A}) are controlled by the
dimensionless expansion parameter $\alpha N^{1/3}/(d\lambda )^{2/3}$.
Inserting the values of the coefficients for $f=1/2$ gives%
\begin{equation}
\left( \chi N\right) _{t}=10.495+41.018\,N^{-1/3}-0.29705\,\alpha ,
\label{chi-t1}
\end{equation}%
\begin{equation}
A_{t}=0.81469\,N^{-1/6}-0.0071843\,N^{1/6}\,\alpha \,.  \label{A2-1}
\end{equation}%
Note that for $N\rightarrow \infty $\ the strength of the electric field $%
E_{0}$\ should tend to zero\ in order that the dimensionless expansion
parameter $\alpha N^{1/3}/(d\lambda )^{2/3}$\ remains small.

Eqs.~(\ref{chi-t1}-\ref{A2-1}) are our main results. The terms in (\ref%
{chi-t1}-\ref{A2-1}) depending on $\alpha $ describe the effects of the
electric field on fluctuations, and were not considered in previous studies
\cite{AHQS93}-\cite{AHQH+S94}. This influence of the electric field on
fluctuations originates from the term $\tilde{\gamma}_{2}^{el}(q)$ under the
integral in Eq.~(\ref{GAns}). Eq.~(\ref{chi-t1}) shows that the electric
field shifts the parameter $\left( \chi N\right) _{t}$ to lower values, and
correspondingly the transition temperature to higher values towards its
mean-field value. The latter means also that the electric field favors
demixing with respect to the free field case. According to Eq.~(\ref{A2-1})
the electric field lowers the value of the order parameter at the transition
point. In other words, the electric field weakens the fluctuations, and
consequently the first order phase transition.

We now will discuss the limits of the Brazovskii-Hartree approach in the
presence of the electric field. The general conditions on the validity of
the Brazovskii-Hartree approach, which are discussed in \cite{FrHe87} \cite%
{swift-hohenberg77}, hold also in the presence of the electric field.
Essentially, while taking into account the fluctuations the peak of the
structure factor at the transition should remain sufficiently sharp, i.e.
the transition should be a weak first-order transition. This requires large
values of N \cite{FrHe87}. A specific approximation used in the presence of
the electric field consists in adopting the expansion of the dielectric
constant in powers of the order parameter to 2nd order, which however is
justified in the vicinity of the transition. The smallness of the linear
term in powers of $\alpha $\ in Eqs.~(\ref{chi-t1}-\ref{A2-1}) imposes a
condition on $E_{0}$: $E_{0}^{2}\ll \lambda ^{2/3}\rho _{m}k_{B}T\varepsilon
_{D}/\beta ^{2}N^{4/3}$. The applicability of the approach of linear
dielectric requires also a limitation on the strength of the electric field.

The dependence of the propagator on $\alpha $ in the field theory associated
with the effective Hamiltonian, which is due to the term $\tilde{\gamma}%
_{2}^{el}(q)$, enables us to make a general conclusion that fluctuations (in
Brazovskii-Hartree approach and beyond) are suppressed for large $\alpha $.
According to this one expects that the order-disorder transition will become
second order for strong fields. The numerical solution of Eqs.~(\ref{T_c})
yields that the mean-field behavior is recovered only in the limit $\alpha
\rightarrow \infty $, which is therefore outside the applicability of linear
electrodynamics. The angular dependence of $\gamma _{4}$, which has not been
taken into account in the present work, gives rise to corrections which are
beyond the linear order of $\alpha $.

The main prediction of the present work that the electric field weakens
fluctuations agrees qualitatively with the behavior of diblock copolymer
melts in shear flow studied in \cite{cates-milner89}, where the shear also
suppresses the fluctuations. Due to the completely different couplings to
the order parameter the calculation schemes are different in both cases.

We now will estimate the shift of the critical temperature for the diblock
copolymer Poly(styrene-block-methylmethacrylate) in an electric field.
Without an electric field its transition temperature is at $182^{\circ }$C
for a molecular weight of $31000$~g/mol~\cite{AHPQ+S92}. We use the
following values of the parameters~\cite{AHQH+S94}: $\varepsilon _{\text{PS}%
}=2.5$, $\varepsilon _{\text{PMMA}}=5.9$\ \footnote{%
these values for $\varepsilon $ refer to the liquid state ($T\approx
160^{\circ }$C), since the corresponding orientation experiments are
performed in this temperature range}, $\beta =\varepsilon _{\text{PMMA}%
}-\varepsilon _{\text{PS}}$, $\varepsilon _{D}=(\varepsilon _{\text{PS}%
}+\varepsilon _{\text{PMMA}})/2$, and $\chi =0.012+17.1/T$. The estimation
of the number of statistical segments $N$\ using the relation $\left( \chi
N\right) _{t}=10.495+41.018\,N^{-1/3}$\ for $T_{t}=182^{\circ }C$\ yields $%
N\approx331$. As described in \cite{AHQH+S94} and \cite{RHS90} for this
calculation a mean statistical segment length $b=7.1~\mathrm{\mathring{A}}$
was assumed, while the approximation $b=\rho _{m}^{-1/3}$\ used here implies
$b\approx 5.2~\mathrm{\mathring{A}}$~\footnote{%
The relation $\rho _{m}=\rho\,N\,N_A/M$, where $N_A$ denotes the Avogadro
number, $M$ the molecular mass of the block copolymer and $\rho\approx1.12~%
\mathrm{g}/\mathrm{cm}^3$ its mass density, results in $\rho_m\approx7.20%
\times10^{21}\mathrm{cm}^{-3}$}. This comparison reflects the limits of the
above approximation. For a field strength $E_{0}=40\mathrm{V/\mu m}$\ the
shift is obtained using Eq.~(\ref{chi-t1}) as%
\begin{equation*}
\Delta T_{t}\approx 2.5\ \mathrm{K}.
\end{equation*}%
The numerical value of the dimensionless expansion parameter, $\alpha
N^{1/3}/(d\lambda )^{2/3}$, is computed in the case under consideration as $%
0.059$. The experimental determination of $\Delta T_{t}$ would be very
helpful to make more detailed fit of the theory to experimental data.

The scattering function in the disordered phase is obtained by taking into
account the composition fluctuations in the presence of the electric field
as
\begin{equation}
S_{dis}(\mathbf{q})\simeq \frac{1}{t_{0}+\alpha \cos ^{2}\theta +(q-q^{\ast
})^{2}}.  \label{S_q}
\end{equation}%
The fluctuational part of \ the scattering function in the ordered state is
obtained as
\begin{equation}
S_{ord}(\mathbf{q})\simeq \frac{1}{t+\alpha \cos ^{2}\theta +(q-q^{\ast
})^{2}}.  \label{Sq-A}
\end{equation}%
At the transition point the expansion of $t_{0}$ and $t$ in powers of the
field strength is derived from Eqs.~(\ref{t}, \ref{t0}) as%
\begin{equation}
t_{0,t}=0.20079(\lambda \,d)^{2/3}\,N^{-1/3}-0.20787\alpha +...,
\label{t0-E}
\end{equation}%
\begin{equation}
t_{t}=1.0591(\lambda \,d)^{2/3}\,N^{-1/3}-0.62147\alpha +....  \label{t-E}
\end{equation}%
The difference between $t_{0,t}$ and $t_{t}$, which is due to the finite
value of the order parameter at the transition, results in the jump of the
peak at the transition point. The structure factor becomes owing to the term
$\alpha \cos ^{2}\theta $ anisotropic in the presence of the electric field.
The structure factor depends on the electric field via $t_{0,t}$ ($t_{t}$)
and the term $\alpha \cos ^{2}\theta $. The suppression of $t_{0,t}$ ($t_{t}$%
) in an electric field according to Eqs.~(\ref{t0-E}, \ref{t-E}) results in
an increase of the peak. Thus, for wave vectors perpendicular to the field
direction, where the angular-dependent term is zero, the peak is more
pronounced than that for $\mathbf{E}_{0}=0$. In the opposite case for wave
vectors parallel to $\mathbf{E}_{0}$ the anisotropic term ($=\alpha \cos
^{2}\theta $) dominates, so that the peak is less pronounced than that for $%
\mathbf{E}_{0}=0$. Composition fluctuations can be associated with
fluctuational modulations of the order parameter. According to Eqs.~(\ref%
{S_q}-\ref{Sq-A}) fluctuational modulations of the order parameter with wave
vectors parallel to the field are strongest suppressed. The latter
correlates with the behavior in the ordered state where the lamellae with
the wave vector perpendicular to the field direction possess the lowest
energy.

\section{Conclusions}

We have generalized the Fredrickson-Helfand theory of microphase separation
in symmetric diblock copolymer melts by taking into account the effects of
the electric field on the composition fluctuations. We have shown that an
electric field suppresses the fluctuations and therefore weakens the
first-order phase transition. However, the mean-field behavior is recovered
in the limit $\alpha \rightarrow \infty $, which is therefore outside the
applicability of the linear electrodynamics. The collective structure factor
in the disordered phase becomes anisotropic in the presence of the electric
field. Fluctuational modulations of the order parameter along the field
direction are strongest suppressed. Thus, the anisotropy of fluctuational
modulations in the disordered state correlates with the parallel orientation
of the lamellae in the ordered state.

\begin{acknowledgments}
\noindent A financial support from the Deutsche Forschungsgemeinschaft, SFB
418 is gratefully acknowledged.
\end{acknowledgments}

\end{document}